\documentclass[prl,aps,showpacs,twocolumn,floatfix]{revtex4}
\usepackage{amsmath,amsfonts,amssymb,graphics,graphicx,color,bm,multirow}
\newcommand{\bS}{{\bm S}}
\newcommand{\scbo}{{\rm SrCu_2\bigl(BO_3\bigr)_2}}
\begin{document}
\title{Local physics of magnetization plateaux in the Shastry-Sutherland model}
\author{L. Isaev$^1$}
\author{G. Ortiz$^1$}
\author{J. Dukelsky$^2$}
\affiliation{$^1$Department of Physics, Indiana University, Bloomington IN
             47405, USA \\
             $^2$Instituto de Estructura de la Materia - CSIC, Serrano 123,
             28006 Madrid, Spain}
\begin{abstract}
 We address the physical mechanism responsible for the emergence of
 magnetization plateaux in the Shastry-Sutherland model. By using a
 hierarchical mean-field approach we demonstrate that a plateau is stabilized
 in a certain {\it spin pattern}, satisfying {\it local} commensurability
 conditions derived from our formalism. Our results provide evidence in favor
 of a robust local physics nature of the plateaux states, and are in agreement
 with recent NMR experiments on $\scbo$.
\end{abstract}
\maketitle

{\it Introduction.--}
The interplay between quantum mechanics and the atomic lattice topology often
leads to a complex mosaic of physical phenomena in low-dimensional frustrated
magnets \cite{Richter}. A prominent representative of this class of materials
is the layered compound $\scbo$, which recently received a lot of attention
because of its fascinating properties in an external magnetic field $h$, namely
the emergence of magnetic plateaux at certain fractions of the saturated
magnetization $M_{\rm sat}$. The first experimental observations of the
plateaux were reported in \cite{Kageyama_1999} for $m=M/M_{\rm sat}=1/8$ and
$1/4$, and somewhat later for $m=1/3$ \cite{Onizuka_2000}. Subsequent nuclear
magnetic resonance (NMR) experiments \cite{Kodama_2002,Takigawa_2006} revealed
spontaneous breaking of the lattice translational symmetry within the $1/8$
plateau, and also indicated that the spin superlattice persists right above
this fraction \cite{Takigawa_2008}. The field was reignited by the work of
Sebastian et al. \cite{Sebastian_2007}, where additional plateaux at exotic
values $m=1/9$, $1/7$, $1/5$ and $2/9$ were reported. However, direct
observation of the emerging spin superstructures remains an experimental
challenge, primarily due to the high magnetic fields ($\sim30-50$ Tesla)
involved in measurements.

\begin{figure}[!t]
 \begin{center}
  \includegraphics[width=0.6\columnwidth]{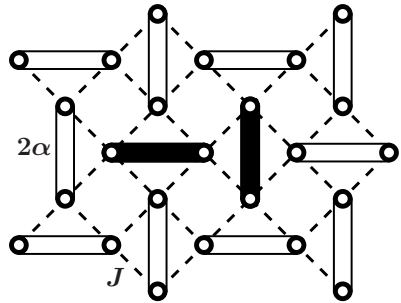}
 \end{center}
 \caption{The SS lattice. Circles denote spins, dashed lines correspond to the
	  NN $J$ coupling, double solid lines denote next-NN interactions
          (dimers). The simplest choice of a degree of freedom, which does not
          cut dimers, is shown in black.}
 \label{fig_tensor}
\end{figure}

The nature of the magnetic states and physical mechanism leading to the
plateaux are also yet to be understood. It is believed that the Heisenberg
antiferromagnetic model on a frustrated Shastry-Sutherland (SS) lattice with
$N$ sites \cite{Shastry_Sutherland} (Fig. \ref{fig_tensor}), 
\begin{equation}
 H=J\sum_{\langle ij\rangle}\bS_i\cdot\bS_j+2J\alpha\sum_{[ij]}\bS_i\cdot\bS_j-
 h\sum_iS^z_i,
 \label{ss_hamiltonian}
\end{equation}
captures the essential magnetic properties of $\scbo$ in relatively high
fields. In Eq. \eqref{ss_hamiltonian} $\bS_i$ denotes a spin-$1/2$
operator at site $i$; the first sum is the usual nearest-neighbor (NN)
Heisenberg term, while the second one runs over dimers; $J$ and
$\alpha\geqslant0$. This model is quasi-exactly solvable
\cite{Shastry_Sutherland} for $\alpha\geqslant1+h/2$: the ground state (GS) is
a direct product of singlet dimer states, and was shown to be stable up to
$\alpha\sim0.71$-$0.75$ in zero field \cite{Miyahara_2003}. In general, it is
an intractable quantum many-body problem where approximation schemes are needed
to deal with large-$N$ systems.

All theories proposed to address this unusual magnetization phenomenon start
from the SS model. However, the physical mechanism stabilizing the plateau
states, their nature, and the structure of the magnetization curve are still
actively debated. Current ideas can be broadly divided into two groups. The
first one advocates subtle non-local (in the spins) correlations leading to an
underlying spin structure which preserves lattice symmetries
\cite{Misguich_2001}. Technically, it employs a mapping of the original spins
to fermions coupled to a Chern-Simons gauge field, and then performs a
Hartree-Fock decoupling. In this way, the qualitative shape of the $\scbo$
magnetization curve was reproduced in high fields, but the lowest plateau at
$1/8$ was missing. Later, this non-local mean-field approach was extended to
include inhomogeneous phases \cite{Sebastian_2007}, and it was argued that the
plateaux correspond to stripe states with broken lattice symmetries.
Remarkably, the length scale $\xi$ associated with the emerging spin
superlattice was found to be $\xi\sim100$ lattice spacings. The second group
contends that the magnetization process can be described in terms of polarized
dimers (triplons), which propagate in the background of singlet dimers
\cite{Miyahara_2003,hc_bosons}. They developed effective hard-core boson models
(truncating the original dimer Hilbert space), solved by perturbative
\cite{Mila_2008} or CORE \cite{Capponi_2008} techniques, and found that the
plateaux states correspond to crystal phases with $\xi\sim10$ lattice
constants.

Such diversity of theoretical predictions demands further investigation. In
this paper we use the hierarchical mean-field (HMF) method \cite{HMF}, in an
attempt to clarify the nature and physical mechanism, responsible for the
emergence of magnetic plateaux in the SS model. Unlike previous calculations,
we deal directly with the SS Hamiltonian \eqref{ss_hamiltonian}, not with 
effective Hamiltonians as in Refs. \cite{Mila_2008,Capponi_2008}, and combine
exact diagonalization data with a simple and {\it controlled} approximation for
the GS wavefunction. For instance, we do not discard the $M=0$ dimer triplet
states, necessary for the propagation of a triplon. We focus on higher-lying
fractions, whose existence has been confirmed experimentally. Our results
support the {\it local} physics nature of the plateau states. In particular,
it is explicitly demonstrated how to construct those robust states based on a
set of commensurability rules that we derived. Our conclusions are also in
agreement with the interpretation of NMR measurements \cite{Kodama_2002}.

{\it Method.--}
The HMF approach is based on the assumption that the physics of the problem is
{\it local} in a particular representation. Since the SS model is formulated in
terms of localized spins, it is natural to work in real space. The main idea of
our method revolves around the concept of a {\it relevant} degree of freedom --
a spin cluster -- which is used to build up the system. Essential quantum
correlations, which drive the physics of the problem, are captured by this
local representation. The SS Hamiltonian is then rewritten in terms of these
coarse-grained variables and a mean-field decoupling is eventually applied to
determine properties of the system. The method is only limited by finite-size
effects and becomes asymptotically exact in the thermodynamic limit. Thus, the
(generally) exponentially hard problem of determining the GS of the model is
reduced to a {\it polynomially} complex one. By identifying each state of a
cluster with a Schwinger boson (SB) and computing matrix elements of spin
operators between these states, one can rewrite {\it exactly} the Hamiltonian
of Eq. \eqref{ss_hamiltonian} in terms of new (cluster) variables
\begin{equation}
 H=\!\!\sum_{i}\epsilon_a(\alpha,h)\gamma^\dag_{ia}\gamma_{ia}+\!\!\sum_{
 \langle ij\rangle_\sigma}\bigl(H^\sigma_{\rm int}\bigr)^{a^\prime
 b^\prime}_{ab}\gamma^\dag_{ia^\prime}\gamma^\dag_{jb^\prime}\gamma_{ia}
 \gamma_{jb}.
 \label{ss_hamiltonian_cluster}
\end{equation}
Here the repeated indices $a$, $b$, etc., which label states of an $N_q$-spin
cluster, are summed over, and $i$ denotes sites in the coarse-grained lattice.
The operators $\gamma_{ia}^\dag$ that create a particular state of a cluster
are $SU\bigl(2^{N_q}\bigr)$ SBs subject to the constraint
$\sum_a\gamma^\dag_{ia} \gamma_{ia}=1$ on each site; $\epsilon_a$ are exact
cluster eigenenergies. Since the original SS Hamiltonian involves only two-spin
interactions, in the new representation there will be only two-boson scattering
processes: the corresponding matrix elements are denoted by
$\bigl(H^\sigma_{\rm int}\bigr)^{a^\prime b^\prime}_{ab}$. The second term in
Eq. \eqref{ss_hamiltonian_cluster} describes the renormalization of the cluster
energy due to its interaction with environment. Thus, our method deals with an
infinite system and finite-size effects enter only through a particular choice
of the cluster. The symbol $\langle ij\rangle_\sigma$, with
$\sigma=1,2,\ldots$, indicates pairs of neighboring blocks, coupled by the same
number of $J$-links.

Application of the HMF method to the SS model starts by recalling that the
phases within plateaux break the lattice translational invariance. Therefore,
the best solution will be obtained, if the degree of freedom matches the unit
cell of the spin superstructure. For each cluster size, $N_q$, and
magnetization
\begin{displaymath}
 m=\frac{1}{M_{\rm sat}}\sum_i\bigl\langle S^z_i\bigr\rangle=
 \frac{2}{N}\sum_i\bigl\langle S^z_i\bigr\rangle=\frac{2}{N_q}
 \sum_{j=1}^{N_q}\bigl\langle S^z_j\bigr\rangle ,
\end{displaymath}
we determine the lowest-energy configuration (i.e., the cluster shape and
corresponding tiling of the lattice). By virtue of previous argument, this
solution will have the ``right'' symmetry. Performing this operation for
successive values of $N_q$ up to the largest one that can be handled, we obtain
a set of magnetization plateaux together with their corresponding spin
profiles. It follows that the particular choice of coarse graining is critical
for the success of this program. One should recall that the experimental value
for $\alpha$ is $0.74$-$0.84$, i.e. the intradimer coupling seems to be ``more
relevant'' than the interdimer one. Therefore, it is natural to consider only
those clusters, which contain an integer number of $\alpha$-links. This
constraint turns out to be quite severe. It follows that the degree of freedom
must also contain an integer number of ``minimal'' blocks, shown in Fig.
\ref{fig_tensor} in black: otherwise the tiling of the lattice will not be
complete. These requirements comprise a set of local commensurability
conditions, necessary to stabilize a plateau.

Another crucial issue is the way the interaction terms in Eq.
\eqref{ss_hamiltonian_cluster} are handled. In an attempt to simplify matters,
we use the straightforward Hartree approximation, i.e., we consider the trial
GS wavefunction
\begin{equation}
 \vert\psi_0\rangle=\prod_i\bigl(R_a\gamma^\dag_{ia}\bigr)\vert0\rangle;\,\,\,
 R^*_aR_a=1.
 \label{GSWF}
\end{equation}
Here $\vert0\rangle$ is the SB vacuum and $R_a$ are variational parameters,
which constitute the cluster wavefunction. Since $H$ is real-valued, we can
choose $R_a$ to be also real. Clearly, this state is cluster translationally
invariant and has exactly one boson per coarse-grained lattice site (so the
constraint is exactly satisfied). Next, we compute the expectation value of $H$
in the state \eqref{GSWF}, subject to periodic boundary conditions, and
minimize it with respect to $R_a$. In this manner one obtains the approximate
GS energy $E_0$ as a function of the magnetic field $h$.

It is important to emphasize the simplicity of our approach. By using a more
sophisticated ansatz (e.g. a Jastrow-type correlated wavefunction \cite{HMF}),
we could improve energies but the physical mechanism and robust structure of
the plateaux will remain intact. Despite its simplicity, the ansatz of Eq.
\eqref{GSWF} was accurate enough to yield the quantitatively correct phase
diagram of the $J_1$-$J_2$ model \cite{HMF}, which involves gapless phases. In
the SS model states within the plateaux are gapped, therefore, our method
should be tailored for this problem.

\begin{figure}[!t]
 \begin{center}
  \includegraphics[width=\columnwidth]{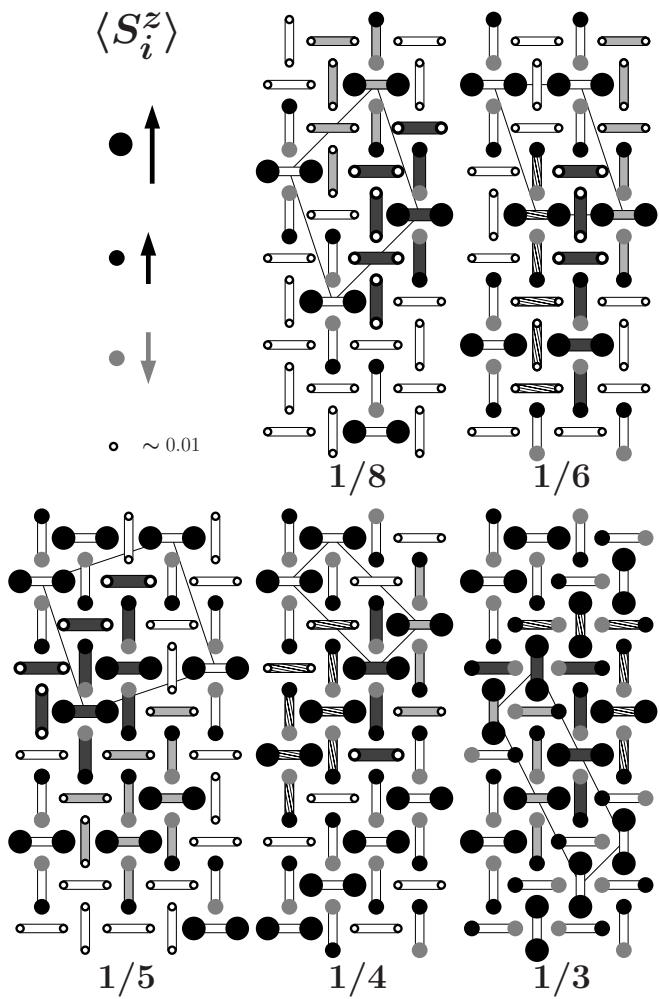}
 \end{center}
 \caption{Schematic spin profiles within plateaux. (Gray) black circles
	  correspond to polarizations (anti) parallel to the field; their
	  sizes encode the magnitude of the local magnetic moment. Empty
	  circles denote sites with $\vert\langle
	  S^z_i\rangle\vert\lesssim10^{-2}$. The clusters used in HMF
	  calculations consist of dark gray dimers. Light gray dimers represent
	  the NN cluster. For $m=1/3$, 1/4 and 1/6 dark and hatched dimers
          constitute the 24-spin cluster. Thin lines indicate unit cells of the
          spin superlattice.}
 \label{fig_local_magnetization}
\end{figure}

{\it Results.--}
To guarantee that our results reduce to the exact solution in the limit
$h\to0$, we mainly consider the region $\alpha\gtrsim 1$. The simplest degree
of freedom, consisting of 4 spins, is shown in Fig. \ref{fig_tensor}. Using
this cluster in our HMF scheme one obtains stable plateaux only at $m=1/2$ and
$m=1$. Clearly, larger blocks are necessary to stabilize plateaux at lower
magnetization fractions. Here, we consider cluster sizes $N_q=4k$ with
$k=2,\ldots,6$ and discuss only plateaux at $1/3$, $1/4$, $1/5$, $1/6$ and
$1/8$, supported in minimal clusters of $N_q=12$, 8, 20, 12 and 16 spins,
respectively ($M_{\rm sat}=N_q/2$). In Fig. \ref{fig_local_magnetization} we
present local spin profiles, corresponding to the lowest energy configurations,
for each of these fractions. Comparison of patterns for different plateaux,
shows that states $1/n$ with $n$ even are characterized by one polarized dimer
per unit cell, while cells of odd-$n$ states have two triplons. For a given
plateau, there typically exist several possible coarse-graining scenarios,
characterized by different clusters and tessellations of the SS lattice, but
identical unit cells. Although these configurations have slightly different
energies, their existence provides an important check for robustness of local
correlations, stabilizing the plateau states. The patterns for all fractions
except 1/5, are similar to those obtained in Refs.
\cite{Kodama_2002,Mila_2008}. Strictly speaking, the profiles in Fig.
\ref{fig_local_magnetization} are well defined only for large values of
$\alpha>1$ and quickly smear out with decreasing $\alpha$. This effect is
difficult to capture within the effective model calculations, like
\cite{Mila_2008}.

\begin{table}[!t]
 \caption{Representative values of the GS energy parameter
          $\varepsilon_0$. Numbers in parenthesis denote the size of
          a cluster, $N_q$.}
 \begin{tabular}{c|c|c|c|c|c}
  $\alpha$ & $1/8$ & $1/6$ & $1/5$ & $1/4$ & $1/3$ \\
  \hline\hline
  \multirow{2}{*}{1.1} & \multirow{2}{*}{0.72056} & 0.68561 (24) &
  \multirow{2}{*}{0.65678} & 0.61291 (24) & 0.53553 (24) \\
  & & 0.68318 (12) & & 0.61121 (16) & 0.53212 (12) \\ 
  \hline
  \multirow{2}{*}{2.0} & \multirow{2}{*}{1.26734} & 1.18978 (24) &
  \multirow{2}{*}{1.12758} & 1.03384 (24) & 0.87689 (24) \\
  & & 1.18937 (12) & & 1.03336 (16) & 0.87569 (12) \\
  \hline\hline
 \end{tabular}
 \label{tab_gse}
\end{table}

\begin{figure}[t]
 \begin{center}
  \includegraphics[width=\columnwidth]{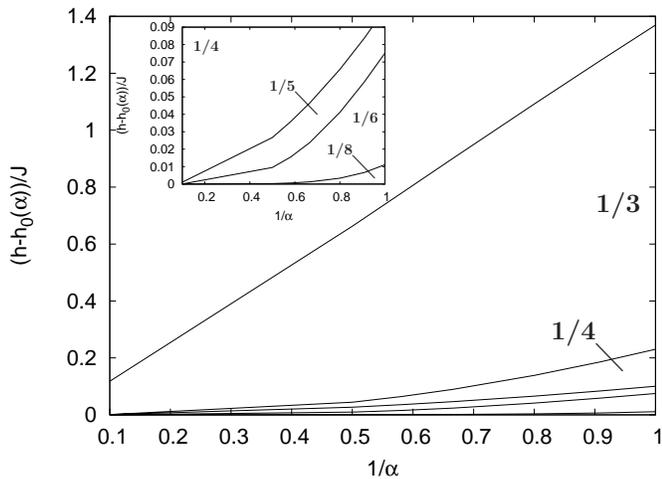}
 \end{center}
 \caption{High magnetic field phase diagram of the SS model for
          $\alpha\geqslant1$. $h_0(\alpha)$ denotes the field after which the
          first plateau (at 1/8) emerges. Fractions indicate values of $m$. For
          $\alpha\gg1$ the triplons
          ({\large$\bullet\!\!\!\!=\!\!\!\!=\!\!\!\!\bullet$})
          in Fig. \ref{fig_local_magnetization} become fully polarized and
          other dimers within the clusters turn into perfect singlets.}
 \label{fig_phase_diagram}
\end{figure}

A clear advantage of our approach, compared to previous works, is its ability
to compute GS energies of the {\it original} SS model. Within each plateau we
have: $E_0(h)/N=-\varepsilon_0-mh/2$. The parameter $\varepsilon_0$ is
presented in Table \ref{tab_gse} for some values of $\alpha$ and different
cluster sizes. In order to address finite-size effects, in Fig.
\ref{fig_phase_diagram} we present the high magnetic field phase diagram of the
SS model for $\alpha\geqslant1$. All fractions were calculated using the
largest possible cluster. Due to the insulating nature of the plateau states,
the finite-size corrections are not expected to significantly affect their
stability. For instance, for $m=1/6$ the energy difference between 12- and
24-spin clusters is only $\sim5\%$ of its width for the values of $\alpha$
shown in Table \ref{tab_gse}. This observation serves as additional evidence in
favor of a universal physical mechanism leading to the plateaux.

As it was already mentioned, the HMF method does not involve truncation of the
dimer Hilbert space. In order to understand consequences of this approximation,
we computed $\varepsilon_0$ for different plateaux, ignoring the dimer state
$\vert\!\!\downarrow\downarrow\rangle$. The resulting absolute error is of the
same order of magnitude as finite size effects and plateau widths (cf. Table
\ref{tab_gse}), which leads to a sizable change in the relative stability of
the plateaux. For example, at $\alpha=1.1$ the average error is
$3\cdot10^{-3}J$, and the lower boundary of the $1/8$ state shifts by $0.04J$.
Therefore, conclusions of the effective boson model calculations, which employ
similar truncation, should generally be taken with caution.

{\it Discussion.--}
Although the effective model approach does yield a sequence of plateaux, their
understanding remains incomplete. Our work addresses this issue by focusing on
the nature and correlations of the magnetic plateau states. In particular, the
analysis presented above, allows the formulation of a set of {\it universal}
rules leading to well-defined {\it spin patterns} (Fig.
\ref{fig_local_magnetization}), which can be probed, e.g. by neutrons. These
rules define a hierarchy of variational plateau wavefunctions and constitute a
central prediction of our work. For a robust state to emerge at a given
magnetization fraction $m$, the {\it commensurability conditions} that have to
be fulfilled are: $(i)$ the (cluster) degree of freedom must contain an even
number of dimers; $(ii)$ the SS lattice must be tessellated completely with
these clusters; $(iii)$ the size of the cluster (unit cell), $N_q$, must allow
the plateau state at $m$, therefore, $N_q=2M/m$ with $M=1,\ldots,N_q/2$ chosen
in a way such that $N_q$ is divisible by four; $(iv)$ the number of triplons
({\huge$\bullet\!\!\!\!=\!\!\!\!=\!\!\!\!\bullet$}) per cluster is $M$ and its
shape must be such that each triplon is surrounded by two dimers of the type
{\large${\color[rgb]{0.7,0.7,0.7}\bullet}\!\!=\!\!\!\!\!=\!\!\!\!\bullet$},
within this cluster. The application of the above constraints leaves us with
the essentially combinatorial problem of actually determining the symmetry and
periodicity of the spin superstructure (see Fig.
\ref{fig_local_magnetization}).

There also exists a number of concrete discrepancies between our work and
recent publications \cite{Mila_2008,Capponi_2008}, which, nevertheless, support
our general conclusion regarding the local nature of the plateau states. First,
the experimentally observed plateaux at $1/4$ and $1/8$, which we found to be
quite robust, were claimed to be unstable in \cite{Mila_2008}. However, the
magnetization profile, presented in Fig. \ref{fig_local_magnetization} for
$m=1/8$, which persists at $\alpha=0.787$, adequate for $\scbo$, is consistent
with the interpretation of available NMR data \cite{Kodama_2002,Takigawa_2006}
for this material. We believe that the origin of these states is purely
magnetic and no additional interactions beyond the SS model are required, in
contradiction with the claim of Ref. \cite{Mila_2008}. Our results also yield a
stable $1/5$ plateau, contrary to the conclusions of Refs.
\cite{Mila_2008,Capponi_2008}. We note that this fraction was observed in
torque measurements of \cite{Sebastian_2007}, however, their proposed spin
superlattice differs dramatically from the one predicted in our Fig.
\ref{fig_local_magnetization}. Another distinction concerns the robustness of
the $1/6$ plateau advocated in \cite{Mila_2008}, which, although present in our
calculation, has a significantly smaller relative width (see the discussion
above). Other fractions at $1/9$, $2/9$ and $2/15$, observed in Refs.
\cite{Mila_2008} and \cite{Capponi_2008}, can also be obtained within our
approach, but this requires significantly larger clusters than the ones used
here. By virtue of our commensurability arguments, we expect the plateaux at
$1/9$ and $2/9$ to emerge in degrees of freedom containing at least 36 spins,
while the $2/15$ fraction will be stabilized in a 60-spin cluster.

Finally we note that the precise shape of the magnetization curve (the relative
energy stability of different plateaux) is quite sensitive to the value of
$\alpha$ (i.e., the particular compound) and, most importantly, since there is
no exact solution of the SS model at these high fields, it depends on the
particular approximation scheme. Experimentally, other physical interactions
not included in the SS model may also add to this uncertainty.

{\it Acknowledgements.--}
We thank C. D. Batista for stimulating discussions. Calculations were performed
on Quarry and NTC clusters at IUB. JD acknowledges support from the Spanish DGI
grant FIS2006-12783-C03-01.

\end{document}